\documentclass[letterpaper,10pt]{scrartcl}
\pdfoutput=1

\usepackage{scrpage2}
\usepackage[USenglish]{babel}

\usepackage{epsfig}

\usepackage{hyperref}
\usepackage[section]{placeins}
\usepackage{amsmath,amssymb,amstext}

\usepackage{prettyref}

\usepackage{subfigure}

\usepackage{dsfont}

\newrefformat{eq}{\textup{(\ref{#1})}}
\newrefformat{lem}{Lemma \ref{#1}}
\newrefformat{thm}{Theorem \ref{#1}}
\newrefformat{cha}{Chapter \ref{#1}}
\newrefformat{sec}{Section \ref{#1}}
\newrefformat{tab}{Table \ref{#1} on page \pageref{#1}}
\newrefformat{fig}{Figure \ref{#1}}

\usepackage{changepage}

\begin{document}

\title{Free energy for parameterized \vskip2mm
Polyakov loops in SU(2) and SU(3)\vskip3mm
 lattice gauge theory}

\begin{titlepage}
\thispagestyle{empty}
$\;$ \hfill INT PUB-12-025
\vskip20mm
  \begin{adjustwidth}{+1cm}{+1cm}
	{\centering
	\makeatletter
	{\sectfont\LARGE \@title }\\
	\vspace{15mm}
	\large{\sf Dmitri Diakonov$^a$, Christof Gattringer$^{b,c}$, Hans-Peter Schadler$^b$} \\
	\vspace{10mm}
	\large{\sf $^a$ Petersburg Nuclear Physics Institute} \\
	\large{\sf Gatchina 188300, St. Petersburg, Russia} \\
	\vspace{5mm}
	\large{\sf $^b$ Institute for Physics, Karl-Franzens-Universit\"at Graz} \\
	\large{\sf Universit\"atsplatz 5, 8010 Graz, Austria} \\
	\vspace{5mm}
	\large{\sf $^c$ Institute for Nuclear Theory, University of Washington} \\
	\large{\sf Box 351560, Seattle, WA 98195-1560, USA} \\
	\vskip30mm
	{\bf Abstract} \\} 
	We present a study of the free energy of parameterized Polyakov loops $P$ in SU($2$) and SU($3$) lattice gauge theory as a function of the parameters that characterize $P$. 
	We explore temperatures below and above the deconfinement transition, and for our highest temperatures 
	$T > 5 \, T_c$ we compare the free energy to perturbative results.
	\vspace{4em}
	\makeatother
  \end{adjustwidth}
\end{titlepage}

\setcounter{page}{1}

% ---------------------------------------------------------------------------
% --------------------- BEGIN Introduction ----------------------------------
% ---------------------------------------------------------------------------
\section{Introduction}
A complete understanding of confinement and the temperature driven transition to a deconfined phase is an important issue of many studies of QCD. While for the full theory even the proper 
characterization of a deconfined phase is under debate, for pure gauge theory the situation is clearer. The transition into the deconfined phase may be related to the spontaneous breaking of 
center symmetry \cite{center}, and a
suitable order parameter is the Polyakov loop $P$, a gauge transporter that closes around the periodic time direction. For temperatures above the 
deconfinement transition $P$ acquires a non-zero expectation value, which after suitable renormalization may be related to the free energy of an isolated static color charge.
An interesting question is to analyze the free energy of the Polyakov loop as a function of the temperature and to monitor its change across the transition. Various approaches to study this problem in an ab-initio lattice setting can be found in the literature \cite{latcenter1,latcenter2}.

In this paper we use a parameterized Polyakov loop as a source and formulate the evaluation of the corresponding free energy on the lattice.
Using Monte Carlo simulation of pure SU(2) and SU(3) lattice gauge theory we study the free energy of the Polyakov loop for a wide range of temperatures and analyze its dependence on the 
parameters characterizing $P$. We monitor the changes at the deconfinement transition and  for the highest temperatures we attempt a comparison to  the perturbative results 
\cite{pert1,pert2,pert3,latpert}.

% ---------------------------------------------------------------------------
% ------- BEGIN Perturbative free energy in the continuum ----------
% ---------------------------------------------------------------------------
\section{Perturbative free energy in the continuum}
In this section we provide a short summary of the perturbative results for the free energy of Polyakov loops at high temperatures  
\cite{pert1,pert2,pert3}. For gauge group SU(N) 
we consider the following parameterization of a diagonal group element:

\begin{equation}
	X\;=\;{\rm diag}\left(e^{i 2\pi  \varphi_1},e^{i 2\pi \varphi_2},\ldots ,e^{i 2\pi \varphi_N}\right)\;,\qquad \varphi_1\!+\!\ldots\!+\varphi_N \; = \; 0 \; .
	\label{defX}
\end{equation}
The parameters $\varphi_j$ are ordered according to 
$\varphi_1\leq\varphi_2\leq\ldots\leq\varphi_N\leq {\varphi_1\!+\!1}$. We now consider the free energy ${\cal F} = - T \log{\cal Z}$ of pure SU(N) gauge theory with the constraint that 
the eigenvalues of the Polyakov loop $P$ assume values
as parameterized above.
By ${\cal Z}$ we denote the partition sum and $T$ is the temperature. We are interested in the free energy ${\cal F}(T,X)$ as a function of the parameterized
SU(N) matrix $X$ and the temperature $T$. 

The free energy  ${\cal F}(T,X)$ can be computed perturbatively  for high temperatures
\cite{pert1,pert2,pert3}. 
Defining the free energy density $F = {\cal F}/V_3$, where $V_3$ is the 3-volume, one can write the one-loop perturbative result as
\begin{equation}
	F(T,X) \;=\;F_0 (T) \; +\; T\,P^{\rm pert}(T,X) \; + \; \ldots \; ,
	\label{FF0}
\end{equation}
where $F_0(T)$ is the result for free thermal gluons, i.e., for a gas of $N^2 - 1$ species of non-interacting vector particles,
\begin{equation}
	F_0(T) \;=\;-\frac{\pi^2}{45}\,T^4\,(N^2-1) \; ,
\end{equation}
and $P^{\rm pert}(T,X)$ is the one-loop contribution to the free energy, which in SU(N) gluodynamics is given by
\begin{equation}
	P^{\rm pert}(T,X) \;=\;\frac{(2\pi)^2T^3}{3}
	\sum_{m>n}^N\left(\varphi_m-\varphi_n\right)^2\left[1-\left(\varphi_m-\varphi_n\right)\right]^2 \; ,
	\label{pertpart}
\end{equation}
where the $\varphi_j$ are the parameters of $X$ as introduced in (\ref{defX}). 
Here we focus on the two cases SU($2$) and SU($3$). For SU($2$) we can write the parameterized Polyakov loop as
\begin{equation}
	X_{\rm SU(2)}\;=\;{\rm diag}\left(e^{i 2\pi  \varphi},e^{-i2\pi \varphi}\right)\quad \mbox{with} \quad 
	-\frac{1}{2}\;\leq \;\varphi \;\leq \;0 \; .
\end{equation}
Introducing $\nu \;=\;-2\varphi, \; \nu \in [0,1]$, the SU(2) perturbative part  reads
\begin{equation}
	P^{\rm pert}_{\rm SU(2)}(T,X)\;=\;\frac{(2\pi)^2T^3}{3}\nu^2(1-\nu)^2,\qquad \nu \;=\;-2\varphi\;,\qquad \nu\in[0,1] \; .
	\label{pertpartSU2}
\end{equation}
For SU($3$) the parameterization for $X$ is given by
\begin{equation}
	X_{\rm SU(3)}\;=\;{\rm diag}\left(e^{i 2\pi  \varphi_1},e^{i 2\pi \varphi_2},
	e^{-i 2\pi (\varphi_1+\varphi_2)}\right)\;,\quad \varphi_1\;\leq\; \varphi_2\;\leq\; -\varphi_1-\varphi_2\;\leq \; 1+\varphi_1 \; ,
	\label{su3param}
\end{equation}
and the perturbative part (\ref{pertpart}) becomes
\begin{align}
	P^{\rm pert}_{\rm SU(3)}(T,X)\;=\;\frac{(2\pi)^2T^3}{3}&\left[ \left(\varphi_1 - \varphi_2\right)^2\left(1+\varphi_1-\varphi_2\right)^2 + \left(\varphi_1+2\varphi_2\right)^2\left(1+\varphi_1+2\varphi_2\right)^2 \right . \notag\\
	&\left. +\left(1+2\varphi_1+\varphi_2\right)^2\left(2\varphi_1+\varphi_2\right)^2  \right]  \; .
	\label{pertpartSU3}
\end{align}
This result can be written as \cite{pert2}
\begin{eqnarray}
	&& P^{\rm pert}_{\rm SU(3)}(T,X)\;=\;f(\varphi_1-\varphi_2)+f(2\varphi_1+\varphi_2)+f(\varphi_1+2\varphi_2) \; ,
\\
&& \mbox{where} \qquad
	f(x)\;=\;\frac{(2\pi)^2T^3}{3}x^2_{\text{mod}\,1}(1-x_{\text{mod}\,1})^2 \; .
\end{eqnarray}
This form now explicitly displays the periodicity of the Polyakov loop parameterization and we can abandon the constraints on the parameters $\varphi_1,\varphi_2$ in (\ref{su3param}).

It is expected that for sufficiently large temperatures the one loop perturbative result describes the free energy (the effective potential) of the parameterized Polyakov loop $P = $ Tr $X$. 
For the case of SU(2) the free energy was worked out also in lattice perturbation theory \cite{latpert}, and we will compare our numerical SU(2) data also to the corresponding results. 
We now formulate the problem on the lattice to study $F(T,X)$ non-perturbatively and also at low temperatures near and below the
deconfinement temperature $T_c$, to analyze whether a qualitative change of  $F(T,X)$ takes place at $T_c$.

% ---------------------------------------------------------------------------
% --------------------- BEGIN Lattice approach ------------------------------
% ---------------------------------------------------------------------------
\section{Formulation on the lattice}
\label{direct}

\subsection{Evaluation of the free energy}
Our lattice formulation of the effective potential is based on the Wilson gauge action 
\begin{equation}\label{eq:wil}
	S[U]  \; = \; \sum_{x,\mu<\nu}\left[1-\frac{1}{N}\text{Re}\,\text{Tr}\,U_{\mu\nu}(x)\right] \; ,
\end{equation}
where the plaquettes $U_{\mu \nu}(x)$ are given by $U_{\mu \nu}(x) \; = \; U_\mu(x) \, U_\nu(x + \hat{\mu}) \,
	U_\mu(x + \hat{\nu})^\dagger \, U_\nu(x)^\dagger$.
The link variables $U_\mu(x)$ are elements of SU(N) attached to the sites of a $N_s^3 \times N_t$ lattice with periodic boundary conditions 
for all directions. $N_s$ is the spatial lattice extent and $N_t$ the number of lattice points in time direction, which is related to the temperature in lattice units via 
$T = 1/N_t$. 
We use the freedom to gauge link variables on subsets of links that do not contain closed contours to arbitrary values, and set all temporal link variables 
to  $\mathds{1}$ except for the first time-slice. The temporal gauge links on that first time-slice are replaced by the parameterized SU(N) element
$X$ defined in (\ref{defX}). We thus work with the following fixed values of the temporal link variables,
\begin{eqnarray}
	U_4(\vec{x},t) & = & \mathds{1} \; \qquad \forall \; \vec{x} \; \;
	\mbox{and} \; \;t = 2, \, \ldots \, N_t \; ,
	\nonumber \\
	U_4(\vec{x},1) & = & X  \qquad  \forall \; \vec{x} \; .
\end{eqnarray}
The spatial link variables remain unrestricted SU(N) matrices and the parameterized matrices $X$ may be viewed as sources for probing the system. The partition sum is 
\begin{equation}
	{\cal Z}(T,X,\beta) \; = \; \int D[U] \, e^{- \beta S[U]} \; ,
	\label{defZ}
\end{equation}
where the path integral $\int\! D[U]$ is only over the spatial gauge variables.  By $\beta$ we denote the inverse gauge coupling which sets the scale, i.e., the lattice spacing we work at. Here we will keep $\beta$ fixed to be able to work at a fixed lattice resolution and drive the temperature $T = 1/N_t$ (in lattice units) by changing the temporal extent $N_t$. It is obvious that with the construction outlined here the Polyakov loop $P$ is given by Tr $X$,
\begin{equation}
P \; = \; \frac{1}{V_3} \sum_{\vec{x}} \mbox{Tr} \, \prod_{t = 1}^{N_t} U_4(\vec{x},t) \; = \; \mbox{Tr} \; X \; ,
\end{equation}
as needed for our analysis.

The partition sum ${\cal Z}$ of Eq.~(\ref{defZ}) depends on the parameterized Polyakov loop $P = $ Tr $X$, 
the temperature $T$ and the inverse gauge coupling $\beta$. By considering its logarithm we can study the free energy as a function of
$X$ and $T$ at a given resolution set by $\beta$.

\begin{figure}[t]
	\centering
	\includegraphics[width=0.75\textwidth]{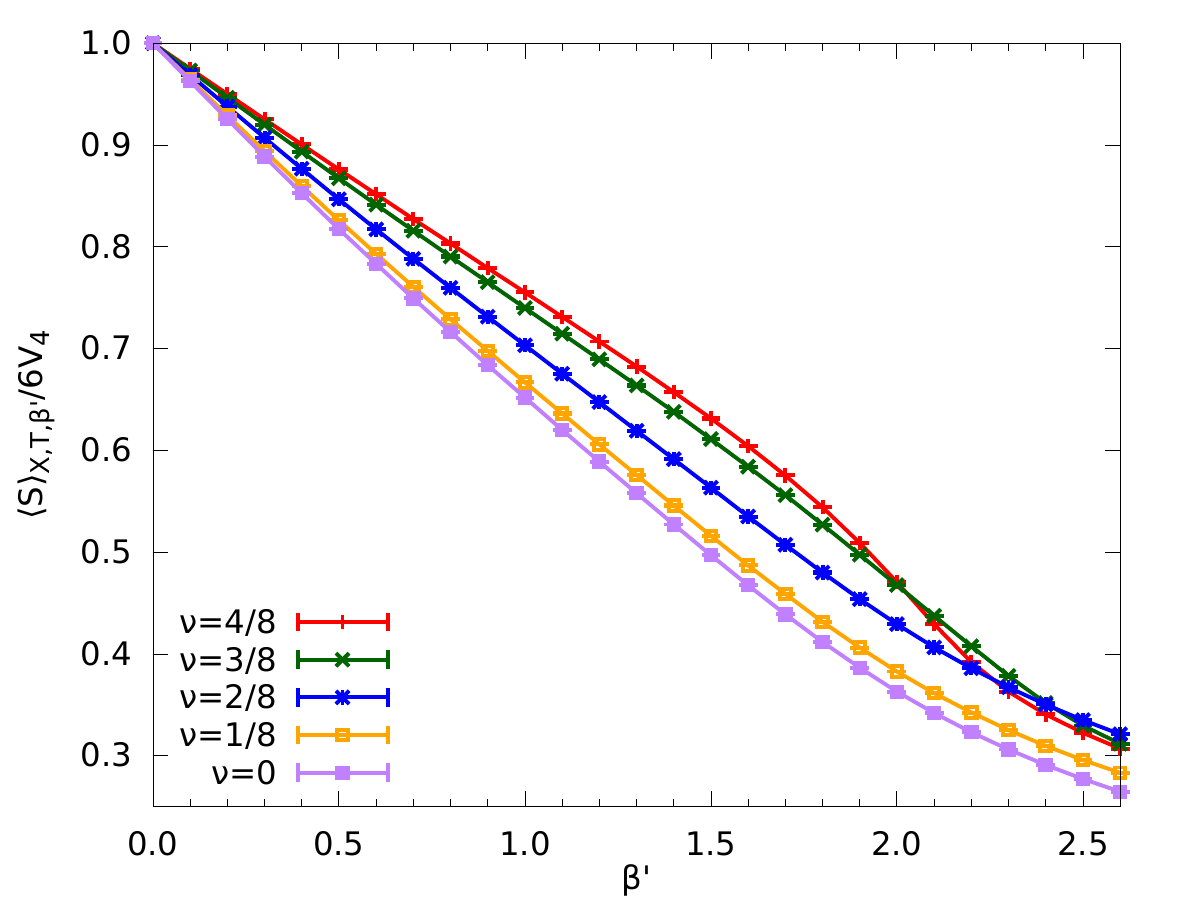}
	\caption{The integrand $\langle S\rangle_{T,X,\beta^\prime}$ of (\protect{\ref{eq:freeint}}) normalized with the number of plaquettes $6 V_4$ as a function of 
	$\beta^\prime$. The example is for SU(2) on  $40^3\times 2$ lattices which corresponds to a temperature of $T = 5.5\,T_c$ and we compare the integrand for 
	different values of the parameter $\nu$.}
	\label{fig:actiont2}
\end{figure}

A technical problem is due to the fact that the partition sum ${\cal Z}(T,X,\beta)$ cannot be computed directly in a Monte Carlo simulation. 
We solve this problem by computing the expectation value 
\begin{equation}
	\langle S\rangle_{T,X,\beta^\prime} \; = \; - \frac{\partial}{\partial \beta^\prime} 
	\ln {\cal Z}(T,X,\beta^\prime)  \; ,
\end{equation}
for several values of $\beta^\prime \leq \beta$ and obtain the free energy ${\cal F}(T,X)$  as
the integral
\begin{equation}\label{eq:freeint}
	{\cal F}(T,X) \; = \; T \int_0^\beta \! d \beta^\prime \, \langle S\rangle_{T,X,\beta^\prime} \; ,
\end{equation}
where the inverse coupling $\beta$ we want to work at (which sets the resolution) appears as the upper limit in the $\beta^\prime$ integral. 
It turns out that the integrand $\langle S\rangle_{T,X,\beta^\prime}$ is very smooth in $\beta^\prime$ (see Fig.~\ref{fig:actiont2} for an SU(2) example) 
such that simple numerical integration techniques allow one to determine the integral (\ref{eq:freeint}) very accurately.

% ---------------------------------------------------------------------------
% --------------------- BEGIN Alternative approach for SU(2) ----------------
% ---------------------------------------------------------------------------

\subsection{Alternative approach for SU(2) with parameter derivatives}
\label{alternative}

For the case of SU(2) we also use an alternative approach which avoids the numerical integral of Eq.~(\ref{eq:freeint}) and needs Monte Carlo 
data at only a single value $\beta$ of the inverse gauge coupling, corresponding to the resolution one wants to work at. The idea is to compute the derivative of 
the free energy density with respect to $\nu$, the single parameter necessary for parameterizing $X$ in the SU(2) case. 
The result then is the vacuum expectation value of a certain functional of the gauge field which can be 
computed with standard Monte Carlo techniques.  A second advantage of the alternative approach is that a possible additive 
constant in the definition of the free energy density is also absent after the $\nu$ derivative.

We start with the necessary derivative of the perturbative solution (\ref{FF0}), (\ref{pertpartSU2}) for the SU(2) case, 
\begin{equation}
	\frac{\partial}{\partial \nu} \, F(T,X)  
	\; = \;   T^4 \, \frac{8\pi^2}{3}\left(2\nu^3-3\nu^2+\nu\right) \; .
	\label{Fderpert}
\end{equation}
On the lattice the corresponding expression is 
\begin{equation}
	\frac{\partial}{\partial \nu} \, F(T,X) \; =  \; -\frac{ T^4 }{ V_3 T^3 }  \, \frac{\partial}{\partial \nu} \, \ln {\cal Z} 	\; = \; 
	\frac{ T \beta }{ V_3 }  \left \langle \frac{\partial S}{\partial \nu}\right \rangle_{T,X,\beta} \; .
	\label{Fderlat}
\end{equation}
In our lattice setting only the temporal links $U_4(\vec{x},1)$ on the first time-slice (they are given by $X$) depend on the parameter $\nu$. Thus for the 
evaluation of $\partial S / \partial \nu$ only the temporal plaquettes at the first time-slice contribute and we find
\begin{eqnarray}
\frac{\partial S}{\partial \nu} &\!\! =\!\! & \frac{\partial}{\partial \nu} \sum_{{\bf x}}\sum_{i=1}^3\left[1-\frac{1}{2}\text{Re}\,\text{Tr}\,U_{i4}({\bf x},t=1)\right] 
\label{Sder}
\\
&\!\! =\!\! &  \frac{\pi}{2}  \text{Im}\! \sum_{x,i} \!\Big(\! \text{Tr}\!\left[U_i({\bf x},1)X(\nu)U_i({\bf x},2)^\dagger Y(\nu)^\dagger\right]
   -  \text{Tr}\!\left[U_i({\bf x},1) Y(\nu)U_i({\bf x},2)^\dagger X(\nu)^\dagger \right]\!\Big)  ,
\nonumber   
\end	{eqnarray}
where 
\begin{equation}
X(\nu) \, = \, \text{diag}\left(e^{-i\pi\nu},e^{i\pi\nu}\right) \quad \mbox{and} \quad
Y(\nu) \, = \, \frac{\partial X(\nu)}{\partial \nu} \, = \, -i\pi\,\text{diag}\left(e^{-i\pi\nu},-e^{i\pi\nu}\right) \; .
\end{equation}
The expression (\ref{Sder}) is a simple functional of the gauge field. Inserting (\ref{Sder}) back into (\ref{Fderlat}) gives the lattice expression 
for $\partial/ \partial \nu \, F(T,X)$ which at high temperatures can be compared to the perturbative result (\ref{Fderpert}).

% ---------------------------------------------------------------------------
% --------------------- BEGIN Results for SU(2) -----------------------------
% ---------------------------------------------------------------------------

\section{Numerical results for SU(2)}

Having discussed the perturbative expectations for $F(T,X)$ at high temperatures and formulated the problem on the lattice we now come to discussing the 
numerical analysis. Our Monte Carlo simulation uses a multi-hit Metropolis update combined with over-relaxation steps \cite{mcover}. 
The results we show for the SU($2$) calculations where obtained on $40^3 \times N_t$ lattices and we varied $N_t = 2,3,\ldots \, 20$ 
to drive the temperature $T = 1/N_t$ (in lattice units). In addition to the runs at $N_s = 40$ we did some finite volume tests with smaller spatial extent which we briefly discuss in the text below.
The statistics is 500 configurations for every set of parameters ($N_t,\, \beta^\prime, \, \nu$). All errors we show in this work are statistical errors determined 
with a jackknife analysis.

For the calculation of $F(T,X)$ with the direct approach based on numerically solving the integral (\ref{eq:freeint}) we set $\beta = 2.6$ for the upper limit of   
the gauge coupling and used 27 sampling points for $\beta^\prime$ between 0 and 2.6 spaced by $\Delta \beta^\prime = 0.1$. 
In a conventional simulation without any prescribed values 
of the temporal links done at $\beta = 2.6$ we scanned the temperature $T = 1/N_t$ by varying $N_t$ between $N_t = 2$ and $N_t = 20$. We evaluated the 
Polyakov loop susceptibility and found its maximum for $N_t = 11$. Thus for $\beta = 2.6$ the deconfinement temperature in lattice units is $T_c \sim 1/11$. 
This implies that when varying $N_t$ between 20 and 2 we cover a range of temperatures from   $T/T_c = 11/20 = 0.55$  at $N_t = 20$ to
$T/T_c = 11/2 = 5.5$  at $N_t = 2$.  

For the parameterization of $X(\nu) = $ diag $( e^{-i\pi\nu},e^{i\pi\nu} )$ we used the values $\nu = n/8$ with $n = 0,1,2,3,4$. For notational convenience, 
in this section we use the notation $F(T,\nu)$ for the free energy density, i.e., in the general notation $F(T,X)$ used so far we 
replace the parameterized matrix $X$ by the single parameter $\nu$ that is 
needed for $X \in $ SU(2). In the plots we show, we explore the symmetries $F(T,\nu) = F(T,-\nu) = F(T,1-\nu)$ to plot our results in the full range of $\nu \in 
[-1,1]$.

\begin{figure}[t]
	\centering
	\includegraphics[width=0.75\textwidth]{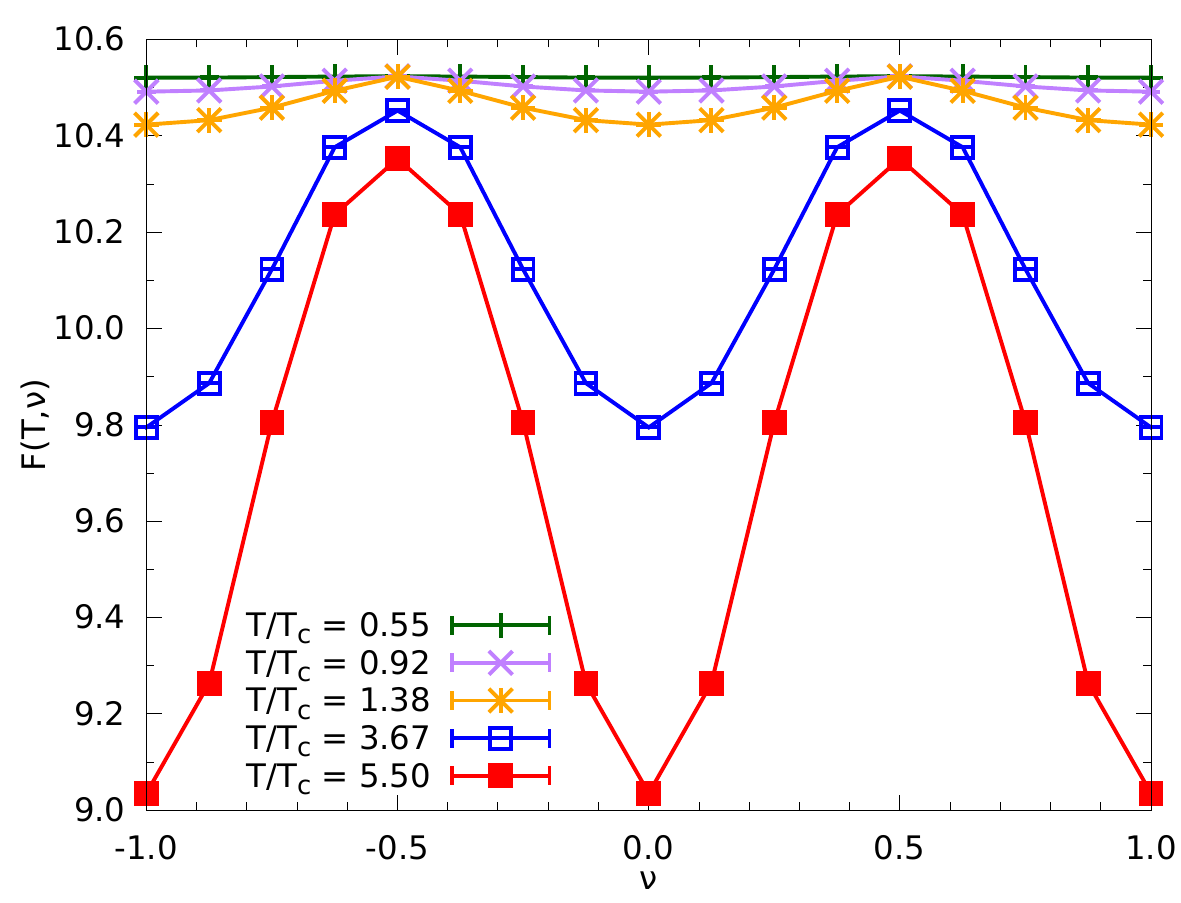}
	\caption{The SU(2) free energy density in lattice units as a function of $\nu$ for different temperatures $T$.}
	\label{fig:fpar}
\end{figure}

We begin the discussion of our results with Fig.~\ref{fig:fpar} where we show the free energy density $F(T,\nu)$ in lattice units as a function of $\nu \in [-1,1]$ and
compare different temperatures $T/T_c$ below and above the deconfinement transition. For the lowest temperature $T = 0.55 \, T_c$ we find that the free energy 
density $F(T,\nu)$ is essentially independent of $\nu$. All values of $\nu$ correspond to the same free energy density and the Polyakov loop
$P = $ Tr $X(\nu)$ averages to zero when $\nu$ fluctuates freely in $[-1,1]$.

As the temperature is increased, one observes the emergence of two minima at $\nu = 0$ and $\nu = 1$ 
(which is equivalent to $\nu = -1$) as one expects from the perturbative high temperature result (\ref{pertpartSU2}). These minima correspond to the two center 
elements $X = \mathds{1}$ and $X = - \mathds{1}$ as is obvious from $X(\nu) = $ diag $( e^{-i\pi\nu},e^{i\pi\nu} )$. With increasing temperature the minima become 
deeper and the breaking of center symmetry corresponds to the Polyakov loop becoming trapped near one of the two center values above $T_c$. 

It is interesting to note that the minima emerge already below the deconfinement temperature $T_c$ and one might suspect that this is due to a finite volume effect. 
To analyze this possibility we repeated the low temperature runs for a smaller spatial volume with $N_s = 30$ and found that the results fall on top of the $N_s = 
40$ data shown in Fig.~\ref{fig:fpar}. We thus can rule out significant effects from finite volume.

\begin{figure}[t]
	\centering
	\includegraphics[width=0.75\columnwidth]{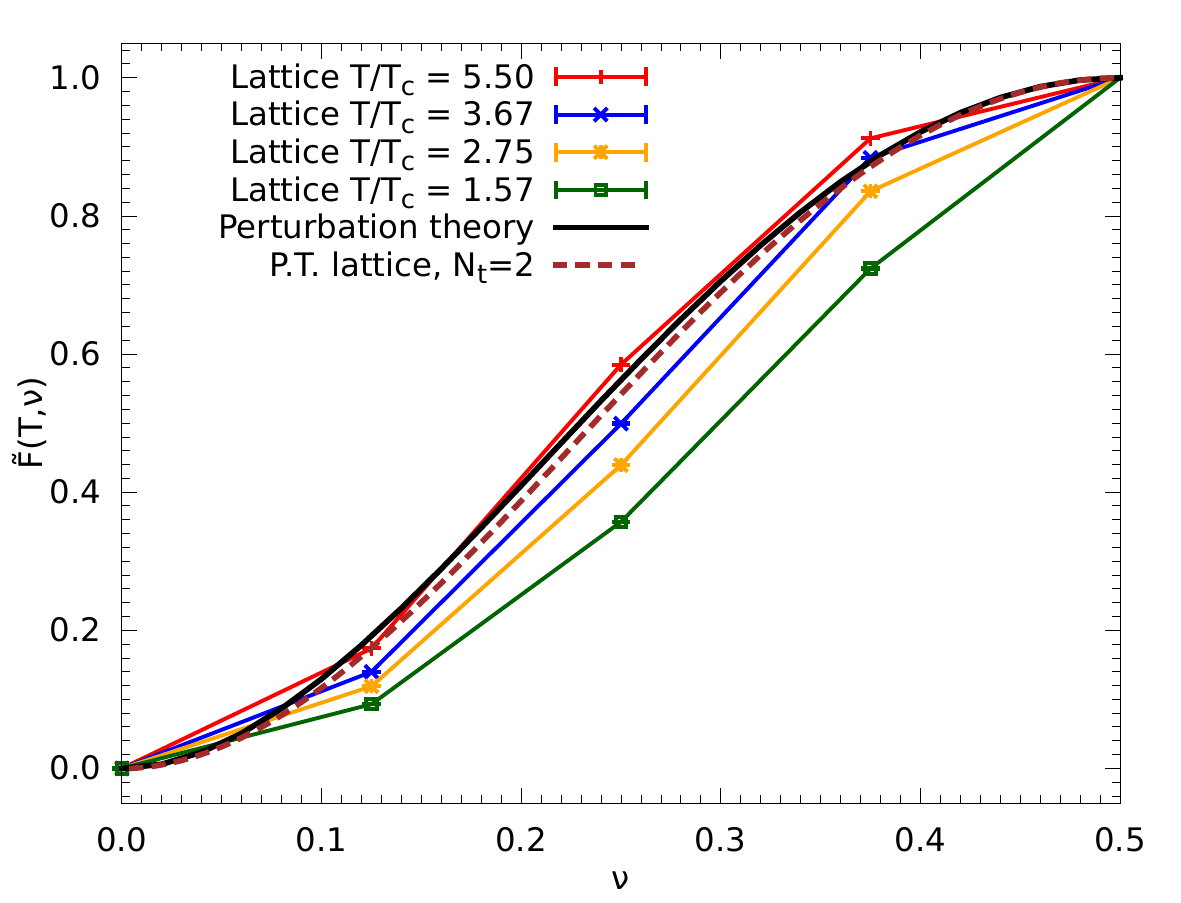}
	\caption{Lattice results (symbols connected with straight lines) for
	the reduced free energy $\widetilde{F}(T,\nu)$ as a function of $\nu$ for several temperatures. The full and dashed curves represent the continuum- and lattice perturbative results. }
	\label{fig:fpol}
\end{figure}

The next step is to compare the lattice results and the perturbative solution. It is important to note that the perturbative continuum setting and our lattice implementation give rise to different additive and multiplicative renormalization of the free energy density $F(T,\nu)$. In order to get rid of this ambiguity and to focus on the $\nu$-dependence,
we consider the reduced free energy density $\widetilde{F}(T,\nu)$ which we define as
\begin{equation}
\widetilde{F}(T,\nu) \; = \; \frac{F(T,\nu) \; - \; F(T,0)}{F(T,1/2) \; - \; F(T,0)} \; .
\label{Fred}
\end{equation}
Subtraction of the value $F(T,0)$ at the minimum removes the additive term, the division by the amplitude $F(T,1/2) \; - \; F(T,0)$ the multiplicative renormalization. A second advantage is that
the reduced free energy $\widetilde{F}(T,\nu)$ is a dimensionless quantity and scale setting ambiguities for the lattice results are avoided. 

In Fig.~\ref{fig:fpol} we show our results for the reduced free energy $\widetilde{F}(T,\nu)$ as a function of $\nu$
at different values of the temperature ranging from $T = 1.57 \, T_c$ to our highest value $T = 5.5 \, T_c$ (full curves). We compare it to the perturbative result
$\widetilde{F}(T,\nu) = 16 \nu^2 (1 - \nu)^2$ (full curve). In addition we display the result from lattice perturbation theory \cite{latpert}  for $N_t = 2$ 
as a dashed curve. The plot illustrates that the lattice data do indeed approach the perturbative 
result with increasing $T$ as expected. The shapes of the continuum and lattice perturbative curves differ only slightly, and it would be interesting to see if higher order  
perturbative results could further improve the agreement with the lattice data.

In addition to the finite volume scaling study we discussed above, we also explored possible discretization effects and compared the results to a cross check done at 
a gauge coupling of $\beta = 2.7$. This can be done at relatively low cost - only one extra point at $\beta = 2.7$ has to be added in the numerical evaluation of the integral (\ref{eq:freeint}). 
We compared the results for $\beta = 2.6$ and 2.7 at values of $N_t$ where the temperatures match and found good agreement.

\begin{figure}[p]
	\centering
	\includegraphics[width=0.7\textwidth]{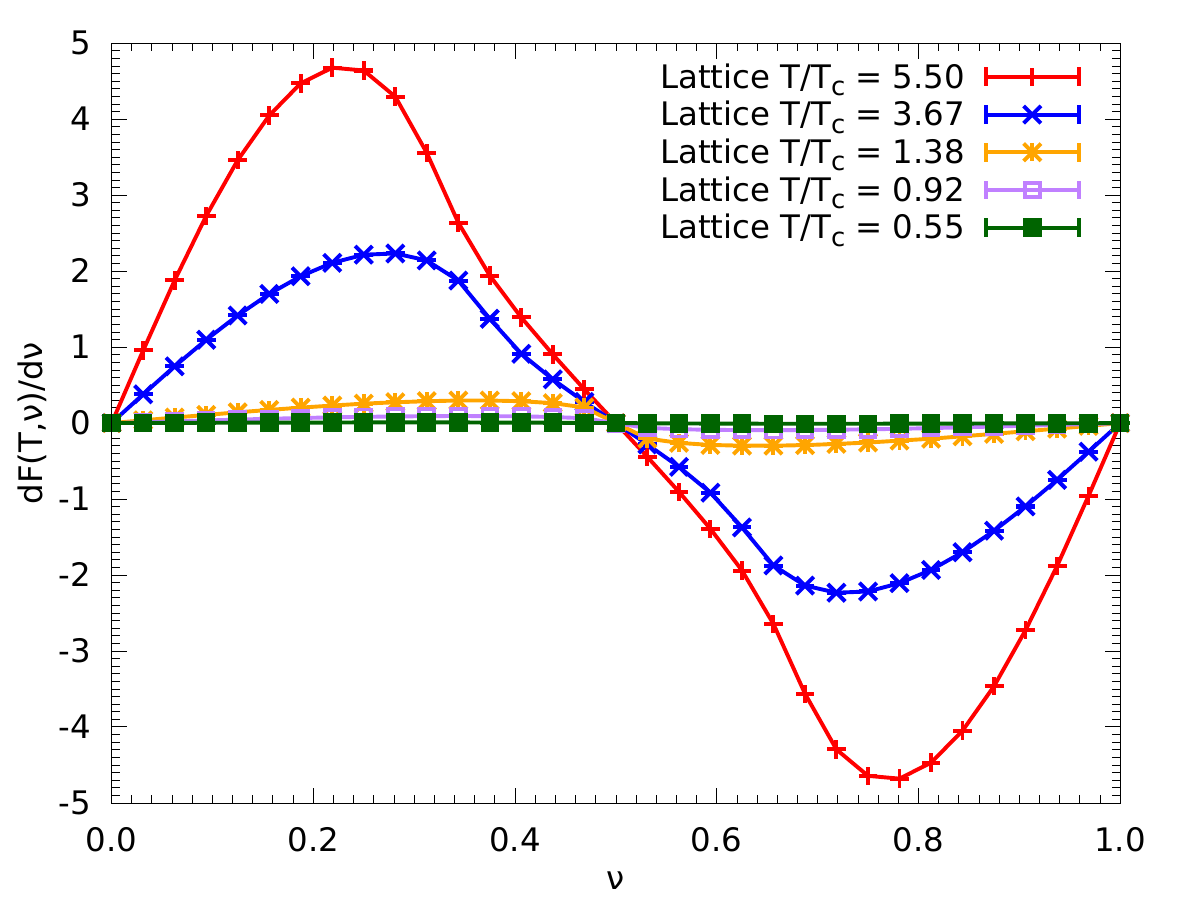}
	\caption{The derivative of the free energy density $\partial / \partial \nu F(T,\nu)$ in lattice units as a function of the parameter $\nu$ for different temperatures below and above $T_c$.}
	\label{fig:derlu}
	\vskip3mm
	\centering
         \includegraphics[width=0.7\textwidth]{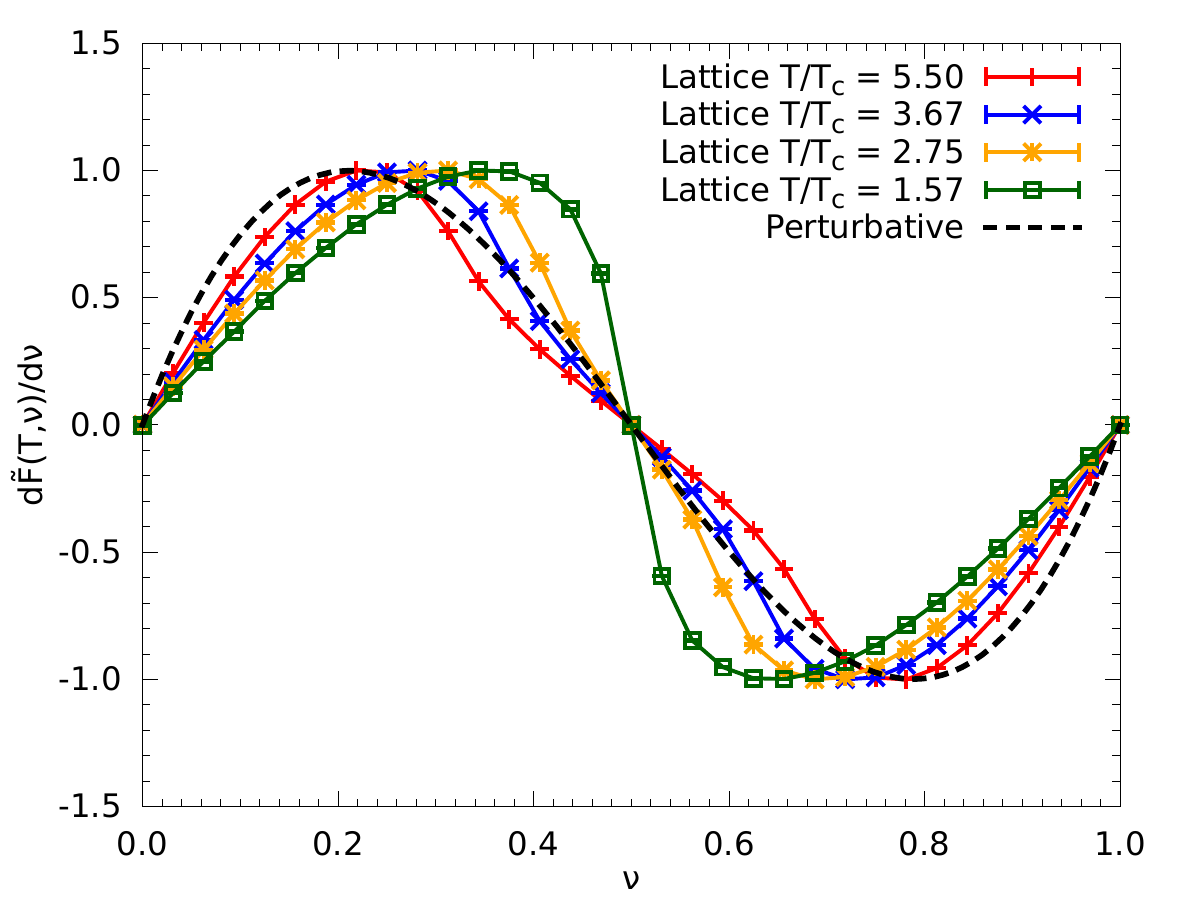}
         \caption{The derivative $\partial / \partial \nu \widetilde{F}(T,\nu)$ versus $\nu$ at various $T > T_c$. 
         The curves are rescaled such that the maxima are normalized to 1.
         We compare the lattice data (symbols connected with straight lines) to the perturbative result (dashed curve).}
	\label{fig:su2dercomp}
\end{figure}

Another consistency test is provided by the alternative approach discussed in Subsection~\ref{alternative}, where the numerical integration 
of  (\ref{eq:freeint}) is avoided (thus for the consistency test we needed only statistics of 100 configurations, but allowed for additional values of $\nu$). 
As for the first approach we begin with an analysis of the results in lattice units. In Fig.~\ref{fig:derlu} we show 
$\partial / \partial \nu F(T,\nu)$ in lattice units as a function of $\nu$  comparing different temperatures. Note that in the alternative approach we can afford many more values of $\nu$
(which we here also show in the larger interval $\nu \in [0,1]$), since the data need to be generated at only a single value of the inverse gauge coupling, i.e., at $\beta = 2.6$.
 Similar to Fig.~\ref{fig:fpar} we find that for our lowest temperatures there is no visible
dependence on the parameter $\nu$. However, already below $T_c$ a $\nu$-dependence sets in, which becomes more and more pronounced at the largest temperatures. As expected, 
$\partial / \partial \nu F(T,\nu)$ vanishes at the values $\nu = 0$ and $\nu = 1/2$ which correspond to the center elements $\pm \mathds{1}$.
 
In Fig.~\ref{fig:su2dercomp} we compare the derivative $\partial / \partial \nu F(T,\nu)$ to the perturbative result (\ref{Fderpert}). Both the lattice data and the perturbative curve are rescaled such that the maxima are normalized to 1, i.e., we again analyze a dimensionless ratio which takes care of the scale setting. We show the results for several values of $T$ and as in the first approach find that the lattice data at our highest temperature $T = 5.5 \, T_c$ are reasonably well described by the one loop perturbative results.

% ---------------------------------------------------------------------------
% --------------------- BEGIN Results for SU(3) -----------------------------
% ---------------------------------------------------------------------------
\section{Numerical results for SU(3)}
Also for the case of SU($3$) the calculations were done on $40^3 \; \times \; N_t$ lattices with $N_t\;=\;2,3,\ldots \, 20$. We worked with an inverse gauge coupling of $\beta = 6.2$ which corresponds to a lattice spacing of $a = 0.068$ fm \cite{Necco} based on the Sommer parameter.  We thus cover a temperature range from 
$T = 146$ MeV at $N_t = 20$ to $T =  1460$ MeV at $N_t = 2$. Using $T_c  \sim  300$ MeV 
this corresponds to a temperature range in units of $T_c$ which is quite similar to the one we covered 
for SU(2). The SU(3) parameterization is based on (\ref{su3param}) with $-1/2 \leq \varphi_1 \leq 0$ and $-1/2 \leq \varphi_2 \leq 0$ for $\varphi_2 \leq \varphi_1$. In particular we evaluate the free energy density for the parameters $\varphi_1 = -k/16 \, , \, k = 0, 1, 2, \ldots , 8$ and $\varphi_2 = -j/16 \, , \, j = 0, 1, 2, \ldots , 8 \, , \, j \geq k$. 
For solving the integral (\ref{eq:freeint}) we now use 
32 sampling values for $\beta^\prime \in [0,6.2]$, spaced by $\Delta \beta^\prime = 0.2$.
 A total of 100 configurations was used for each parameter combination $\varphi_1,\varphi_2, \beta^\prime$.

\begin{figure}[p]
	\centering
	\includegraphics[width=0.75\textwidth]{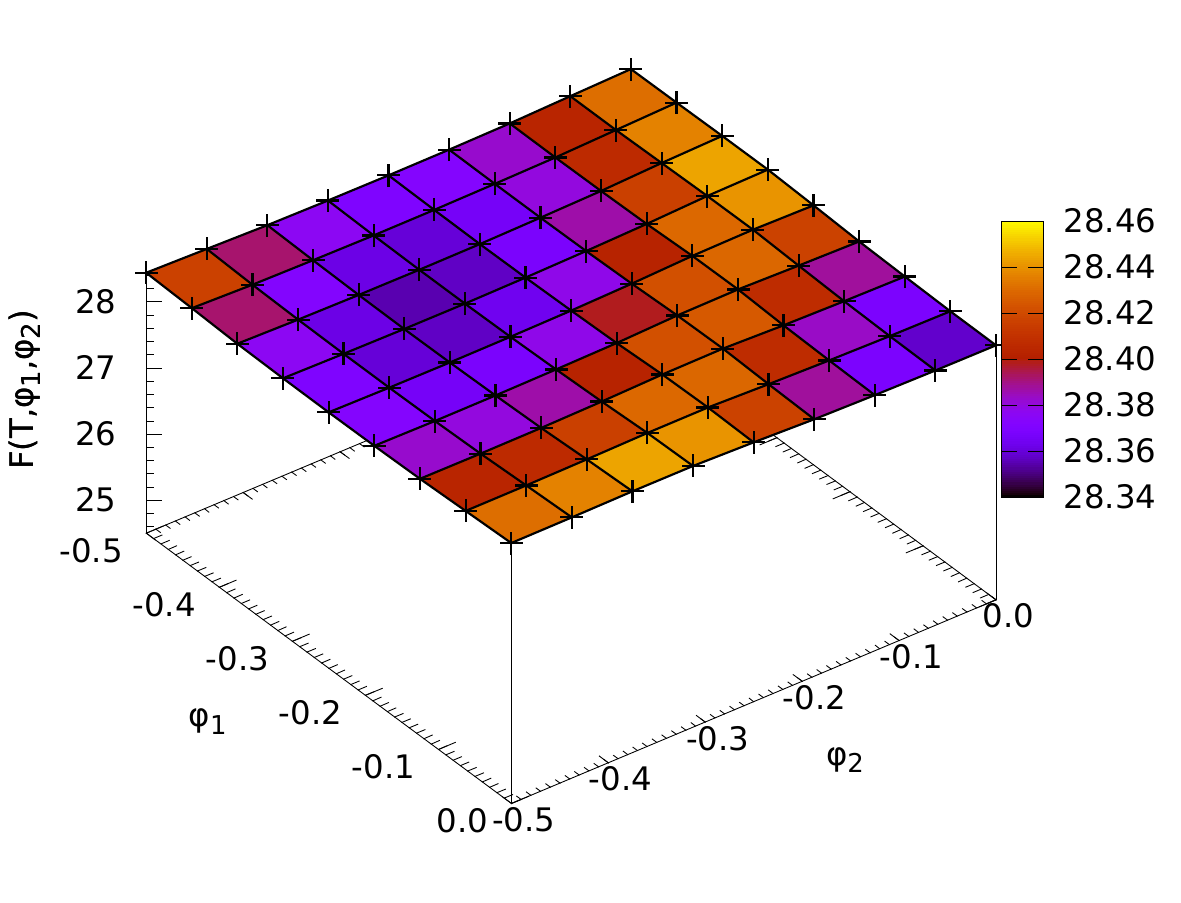}	
	\includegraphics[width=0.75\textwidth]{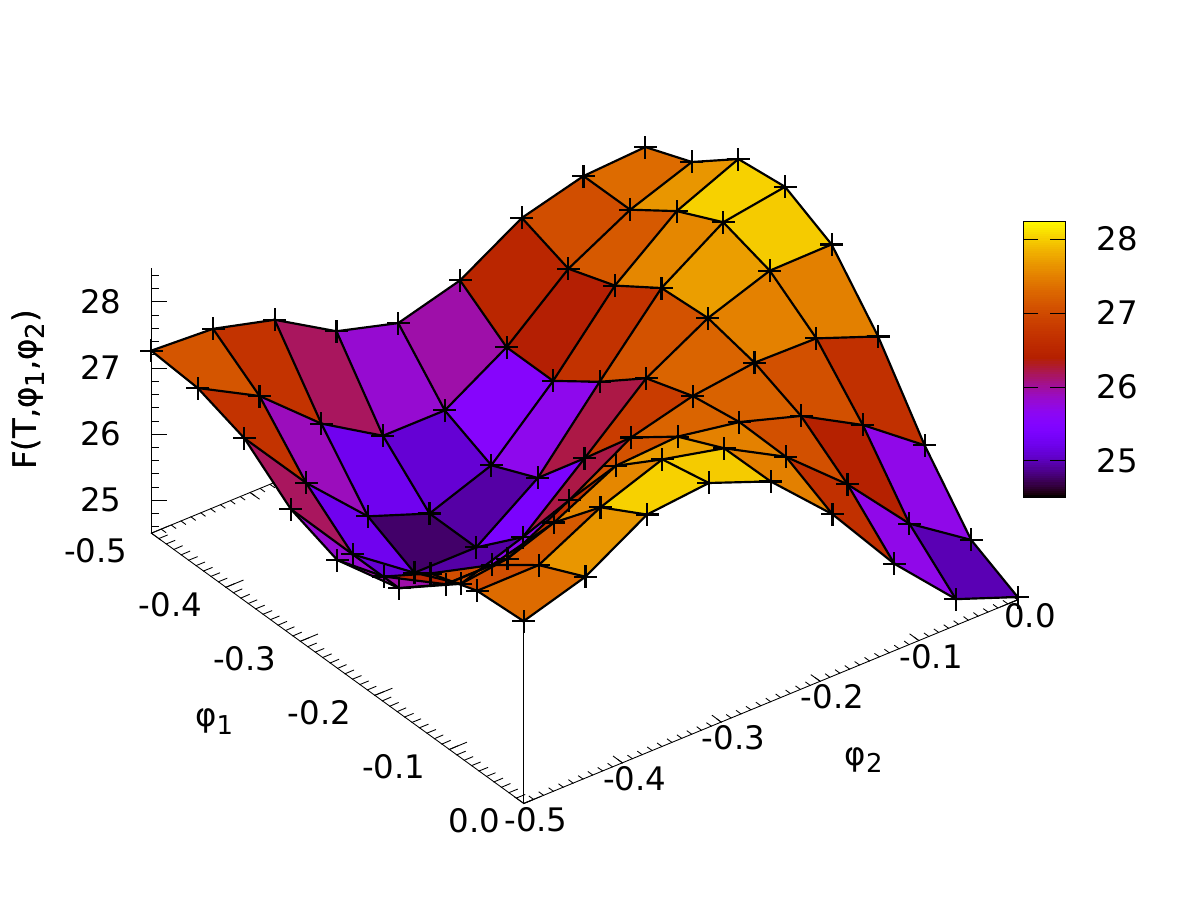}
	\caption{SU(3) Free energy density $F(T,\varphi_1,\varphi_2)$ in lattice units as a function of $\varphi_1, \varphi_2$. We show the results at $T =  290$ MeV (top plot) and at $T = 1460$ MeV  
	(bottom). Note that we use a different color coding for the two temperatures.}
	\label{fig:su3fenergy}
\end{figure}

In Fig.~\ref{fig:su3fenergy} we show 3D plots of the SU(3) free energy density $F(T,\varphi_1,\varphi_2)$ 
in lattice units as a function of $\varphi_{1}$ and $\varphi_2$. We compare the results at $T = 290$ MeV (top plot) and at
$T = 1460$ MeV, and stress again that we work at fixed $\beta = 6.2$, i.e., the lattice units are the same for both plots. At large temperatures one expects to find minima at the three center elements $\mathds{1}$ which corresponds to $\varphi_1 = \varphi_2 = 0$, at 
$e^{i 2\pi/3} \, \mathds{1}$, i.e., $\varphi_1 = \varphi_2 = 1/3$ and at $e^{-i 2\pi/3} \, \mathds{1}$ at $\varphi_1 = \varphi_2 = -1/3$. In the parameter range we use for our plots, the minima at 
$\varphi_1 = \varphi_2 = 0$ and $\varphi_1 = \varphi_2 = -1/3$ are clearly visible in the $T = 1460$ MeV plot (bottom).   For $T = 290$ MeV (top), which is just below the deconfinement temperature,  the minima are essentially gone and are only visible when inspecting the color coding (note that the scale for the color coding is different for the two temperatures).
This situation is different from the SU(2) case where the minima are clearly visible also below $T_c$, and has a natural explanation from the fact that for pure SU(2) gauge theory one has 
 a second order transition, while the transition is of first order for SU(3). 

\begin{figure}[t]
	\centering
         \includegraphics[width=0.9\textwidth,clip]{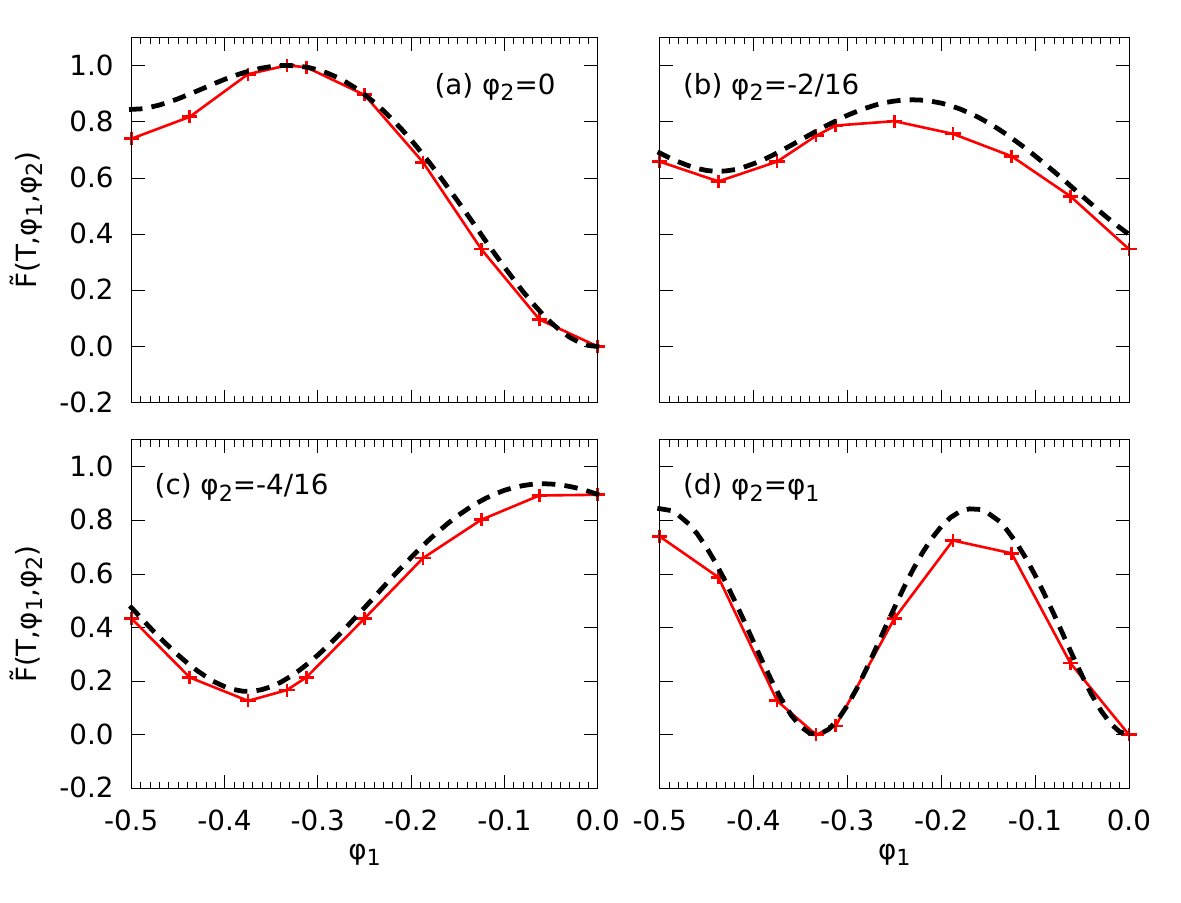}
	\caption{Comparison of SU($3$) perturbative and lattice results for
	the reduced free energy $\widetilde{F}(T,\varphi_1,\varphi_2)$ as a function of $\varphi_1$ for different $\varphi_2$ at our highest  
	temperature $T  \approx 1460 \, \mathrm{MeV}$. The lattice data are represented by symbols connected with straight lines and the perturbative results by dashed curves.}
	\label{fig:su3comp}
\end{figure}

For the comparison with the perturbative SU($3$)  results we again consider the dimensionless reduced free energy $\widetilde{F}(T,\varphi_1,\varphi_2)$ which for SU(3) is defined as
\begin{equation}
\widetilde{F}(T,\varphi_1,\varphi_2) \; = \; \frac{F(T,\varphi_1,\varphi_2) \; - \; F(T,0,0)}{F(T,-1/3,0) \; - \; F(T,0,0)} \; .
\label{Fred2}
\end{equation}
In Fig.~\ref{fig:su3comp} we study sections of  $\widetilde{F}(T,\varphi_1,\varphi_2)$ at $\varphi_2 = 0$,  $\varphi_2 = -  2/16$ and $\varphi_2 = -4/16$, as well as the 
case $\varphi_1 = \varphi_2$ and plot $\widetilde{F}(T,\varphi_1,\varphi_2)$ as a function of $\varphi_1$ for the listed choices of $\varphi_2$. In the plots we show the lattice results with symbols connected with straight lines, and the perturbative results as dashed curves.
 
% ---------------------------------------------------------------------------
% -------------------------- BEGIN Conclusion -------------------------------
% ---------------------------------------------------------------------------
\section{Conclusion}
In this work we have presented a study of the free energy of a parameterized Polyakov loop in SU(2) and SU(3) lattice gauge theory. The Polyakov loop was introduced as a parameterized source and we studied the system using Monte Carlo simulation in a wide range of temperatures below and above the deconfinement transition. We used a fixed scale, i.e., we work at a fixed inverse gauge coupling $\beta$ and drive the temperature by changing the temporal extent $N_t$, such that we do not mix discretization, finite volume and temperature effects. 

For temperatures above the deconfinement transition the free energy of the Polyakov loop develops the expected minima near the parameter values that correspond to the center elements. For the highest temperatures ($T > 5 \, T_c$) we studied a reduced free energy to remove the additive term and to make it dimensionless. In this form the dependence of the free energy was compared to one-loop perturbative results which we found to reasonable describe the lattice data. For our lowest temperature $T \sim 0.5 \, T_c$ we observed that the free energy is independent of the parameters of the Polyakov loop, both for gauge groups SU(2) and SU(3). For the SU(2) case, where the deconfinement transition is of second order, we found that already below $T_c$ the minima at the center elements start to emerge.  For SU(3) with its first order transition the free energy remains independent of the Polyakov loop parameters up to $T_c$.

\vspace{1em}

\noindent
{\bf Acknowledgments:}
We thank Philippe de Forcrand, Christian Lang, Axel Maas and Victor Petrov for discussions.
C. Gattringer thanks the members of the INT at the University of Washington where part of this work was done for their hospitality, and the Dr. Heinrich J\"org Foundation at the 
Karl-Franzens-University Graz for support. 

% References
%\bibliographystyle{h-physrev}
%\bibliography{effectivepotential.bib}

\end{document}